\documentclass{nature}

\usepackage{amssymb}
\usepackage{amsmath}
\usepackage{wasysym}
\usepackage{graphicx}
\usepackage{xspace}
\usepackage[usenames, dvipsnames]{color}
\usepackage[normalem]{ulem}
\usepackage{soul}

\usepackage[displaymath,mathlines]{lineno}
\usepackage{subfig}
\usepackage[font=footnotesize]{caption}

\newcommand{\Fermi}{\emph{Fermi}\xspace}
\newcommand{\Swift}{\emph{Swift}\xspace}
\newcommand{\photonindex}{\gamma}

\usepackage{floatrow}
\DeclareFloatFont{tiny}{\scriptsize}

\title{A very-high-energy component deep in the Gamma-ray Burst afterglow}

\newcounter{firstbib}

\begin{document}

\maketitle

\author{
H.~Abdalla$^{\ref{NWU}}$,
R.~Adam$^{\ref{LLR}}$,
F.~Aharonian$^{\ref{MPIK},\ref{DIAS},\ref{RAU}}$,
F.~Ait~Benkhali$^{\ref{MPIK}}$,
E.O.~Ang\"uner$^{\ref{CPPM}}$,
M.~Arakawa$^{\ref{Rikkyo}}$,
C.~Arcaro$^{\ref{NWU}}$,
C.~Armand$^{\ref{LAPP}}$,
H.~Ashkar$^{\ref{IRFU}}$,
M.~Backes$^{\ref{UNAM},\ref{NWU}}$,
V.~Barbosa~Martins$^{\ref{DESY}}$,
M.~Barnard$^{\ref{NWU}}$,
Y.~Becherini$^{\ref{Linnaeus}}$,
D.~Berge$^{\ref{DESY}}$,
K.~Bernl\"ohr$^{\ref{MPIK}}$,
E.~Bissaldi$^{\ref{INFNBari},\ref{PBari}}$,
R.~Blackwell$^{\ref{Adelaide}}$,
M.~B\"ottcher$^{\ref{NWU}}$,
C.~Boisson$^{\ref{LUTH}}$,
J.~Bolmont$^{\ref{LPNHE}}$,
S.~Bonnefoy$^{\ref{DESY}}$,
J.~Bregeon$^{\ref{LUPM}}$,
M.~Breuhaus$^{\ref{MPIK}}$,
F.~Brun$^{\ref{IRFU}}$,
P.~Brun$^{\ref{IRFU}}$,
M.~Bryan$^{\ref{GRAPPA}}$,
M.~B\"{u}chele$^{\ref{ECAP}}$,
T.~Bulik$^{\ref{UWarsaw}}$,
T.~Bylund$^{\ref{Linnaeus}}$,
M.~Capasso$^{\ref{IAAT}}$,
S.~Caroff$^{\ref{LPNHE}}$,
A.~Carosi$^{\ref{LAPP}}$,
S.~Casanova$^{\ref{IFJPAN},\ref{MPIK}}$,
M.~Cerruti$^{\ref{LPNHE},\ref{CerrutiNowAt}}$,
T.~Chand$^{\ref{NWU}}$,
S.~Chandra$^{\ref{NWU}}$,
A.~Chen$^{\ref{WITS}}$,
S.~Colafrancesco$^{\ref{WITS}}$ \protect\footnotemark[2], 
M.~Cury{\l}o$^{\ref{UWarsaw}}$,
I.D.~Davids$^{\ref{UNAM}}$,
C.~Deil$^{\ref{MPIK}}$,
J.~Devin$^{\ref{CENBG}}$,
P.~deWilt$^{\ref{Adelaide}}$,
L.~Dirson$^{\ref{HH}}$,
A.~Djannati-Ata\"i$^{\ref{APC}}$,
A.~Dmytriiev$^{\ref{LUTH}}$,
A.~Donath$^{\ref{MPIK}}$,
V.~Doroshenko$^{\ref{IAAT}}$,
J.~Dyks$^{\ref{NCAC}}$,
K.~Egberts$^{\ref{UP}}$,
G.~Emery$^{\ref{LPNHE}}$,
J.-P.~Ernenwein$^{\ref{CPPM}}$,
S.~Eschbach$^{\ref{ECAP}}$,
K.~Feijen$^{\ref{Adelaide}}$,
S.~Fegan$^{\ref{LLR}}$,
A.~Fiasson$^{\ref{LAPP}}$,
G.~Fontaine$^{\ref{LLR}}$,
S.~Funk$^{\ref{ECAP}}$,
M.~F\"u{\ss}ling$^{\ref{DESY}}$,
S.~Gabici$^{\ref{APC}}$,
Y.A.~Gallant$^{\ref{LUPM}}$,
F.~Gat{\'e}$^{\ref{LAPP}}$,
G.~Giavitto$^{\ref{DESY}}$,
L.~Giunti$^{\ref{APC}}$,
D.~Glawion$^{\ref{LSW}}$,
J.F.~Glicenstein$^{\ref{IRFU}}$,
D.~Gottschall$^{\ref{IAAT}}$,
M.-H.~Grondin$^{\ref{CENBG}}$,
J.~Hahn$^{\ref{MPIK}}$,
M.~Haupt$^{\ref{DESY}}$,
G.~Heinzelmann$^{\ref{HH}}$,
G.~Henri$^{\ref{Grenoble}}$,
G.~Hermann$^{\ref{MPIK}}$,
J.A.~Hinton$^{\ref{MPIK}}$,
W.~Hofmann$^{\ref{MPIK}}$,
C.~Hoischen$^{\ref{UP}}$,
T.~L.~Holch$^{\ref{HUB}}$,
M.~Holler$^{\ref{LFUI}}$,
D.~Horns$^{\ref{HH}}$,
D.~Huber$^{\ref{LFUI}}$,
H.~Iwasaki$^{\ref{Rikkyo}}$,
M.~Jamrozy$^{\ref{UJK}}$,
D.~Jankowsky$^{\ref{ECAP}}$,
F.~Jankowsky$^{\ref{LSW}}$,
A.~Jardin-Blicq$^{\ref{MPIK}}$,
I.~Jung-Richardt$^{\ref{ECAP}}$,
M.A.~Kastendieck$^{\ref{HH}}$,
K.~Katarzy{\'n}ski$^{\ref{NCUT}}$,
M.~Katsuragawa$^{\ref{KAVLI}}$,
U.~Katz$^{\ref{ECAP}}$,
D.~Khangulyan$^{\ref{Rikkyo}}$,
B.~Kh\'elifi$^{\ref{APC}}$,
J.~King$^{\ref{LSW}}$,
S.~Klepser$^{\ref{DESY}}$,
W.~Klu\'{z}niak$^{\ref{NCAC}}$,
Nu.~Komin$^{\ref{WITS}}$,
K.~Kosack$^{\ref{IRFU}}$,
D.~Kostunin$^{\ref{DESY}}$,
M.~Kreter$^{\ref{NWU}}$,
G.~Lamanna$^{\ref{LAPP}}$,
A.~Lemi\`ere$^{\ref{APC}}$,
M.~Lemoine-Goumard$^{\ref{CENBG}}$,
J.-P.~Lenain$^{\ref{LPNHE}}$,
E.~Leser$^{\ref{UP},\ref{DESY}}$,
C.~Levy$^{\ref{LPNHE}}$,
T.~Lohse$^{\ref{HUB}}$,
I.~Lypova$^{\ref{DESY}}$,
J.~Mackey$^{\ref{DIAS}}$,
J.~Majumdar$^{\ref{DESY}}$,
D.~Malyshev$^{\ref{IAAT}}$,
V.~Marandon$^{\ref{MPIK}}$,
A.~Marcowith$^{\ref{LUPM}}$,
A.~Mares$^{\ref{CENBG}}$,
C.~Mariaud$^{\ref{LLR}}$,
G.~Mart\'i-Devesa$^{\ref{LFUI}}$,
R.~Marx$^{\ref{MPIK}}$,
G.~Maurin$^{\ref{LAPP}}$,
P.J.~Meintjes$^{\ref{UFS}}$,
A.M.W.~Mitchell$^{\ref{MPIK},\ref{MitchellNowAt}}$,
R.~Moderski$^{\ref{NCAC}}$,
M.~Mohamed$^{\ref{LSW}}$,
L.~Mohrmann$^{\ref{ECAP}}$,
C.~Moore$^{\ref{Leicester}}$,
E.~Moulin$^{\ref{IRFU}}$,
J.~Muller$^{\ref{LLR}}$,
T.~Murach$^{\ref{DESY}}$,
S.~Nakashima $^{\ref{RIKKEN}}$,
M.~de~Naurois$^{\ref{LLR}}$,
H.~Ndiyavala $^{\ref{NWU}}$,
F.~Niederwanger$^{\ref{LFUI}}$,
J.~Niemiec$^{\ref{IFJPAN}}$,
L.~Oakes$^{\ref{HUB}}$,
P.~O'Brien$^{\ref{Leicester}}$,
H.~Odaka$^{\ref{Tokyo}}$,
S.~Ohm$^{\ref{DESY}}$,
E.~de~Ona~Wilhelmi$^{\ref{DESY}}$,
M.~Ostrowski$^{\ref{UJK}}$,
I.~Oya$^{\ref{DESY}}$,
M.~Panter$^{\ref{MPIK}}$,
R.D.~Parsons$^{\ref{MPIK}}$,
C.~Perennes$^{\ref{LPNHE}}$,
P.-O.~Petrucci$^{\ref{Grenoble}}$,
B.~Peyaud$^{\ref{IRFU}}$,
Q.~Piel$^{\ref{LAPP}}$,
S.~Pita$^{\ref{APC}}$,
V.~Poireau$^{\ref{LAPP}}$,
A.~Priyana~Noel$^{\ref{UJK}}$,
D.A.~Prokhorov$^{\ref{WITS}}$,
H.~Prokoph$^{\ref{DESY}}$,
G.~P\"uhlhofer$^{\ref{IAAT}}$,
M.~Punch$^{\ref{APC},\ref{Linnaeus}}$,
A.~Quirrenbach$^{\ref{LSW}}$,
S.~Raab$^{\ref{ECAP}}$,
R.~Rauth$^{\ref{LFUI}}$,
A.~Reimer$^{\ref{LFUI}}$,
O.~Reimer$^{\ref{LFUI}}$,
Q.~Remy$^{\ref{LUPM}}$,
M.~Renaud$^{\ref{LUPM}}$,
F.~Rieger$^{\ref{MPIK}}$,
L.~Rinchiuso$^{\ref{IRFU}}$,
C.~Romoli$^{\ref{MPIK}}$,
G.~Rowell$^{\ref{Adelaide}}$,
B.~Rudak$^{\ref{NCAC}}$,
E.~Ruiz-Velasco$^{\ref{MPIK}}$,
V.~Sahakian$^{\ref{YPI}}$,
S.~Sailer$^{\ref{MPIK}}$,
S.~Saito$^{\ref{Rikkyo}}$,
D.A.~Sanchez$^{\ref{LAPP}}$,
A.~Santangelo$^{\ref{IAAT}}$,
M.~Sasaki$^{\ref{ECAP}}$,
R.~Schlickeiser$^{\ref{RUB}}$,
F.~Sch\"ussler$^{\ref{IRFU}}$,
A.~Schulz$^{\ref{DESY}}$,
H.M.~Schutte$^{\ref{NWU}}$,
U.~Schwanke$^{\ref{HUB}}$,
S.~Schwemmer$^{\ref{LSW}}$,
M.~Seglar-Arroyo$^{\ref{IRFU}}$,
M.~Senniappan$^{\ref{Linnaeus}}$,
A.S.~Seyffert$^{\ref{NWU}}$,
N.~Shafi$^{\ref{WITS}}$,
K.~Shiningayamwe$^{\ref{UNAM}}$,
R.~Simoni$^{\ref{GRAPPA}}$,
A.~Sinha$^{\ref{APC}}$,
H.~Sol$^{\ref{LUTH}}$,
A.~Specovius$^{\ref{ECAP}}$,
M.~Spir-Jacob$^{\ref{APC}}$,
{\L.}~Stawarz$^{\ref{UJK}}$,
R.~Steenkamp$^{\ref{UNAM}}$,
C.~Stegmann$^{\ref{UP},\ref{DESY}}$,
C.~Steppa$^{\ref{UP}}$,
T.~Takahashi $^{\ref{KAVLI}}$,
T.~Tavernier$^{\ref{IRFU}}$,
A.M.~Taylor$^{\ref{DESY}}$,
R.~Terrier$^{\ref{APC}}$,
D.~Tiziani$^{\ref{ECAP}}$,
M.~Tluczykont$^{\ref{HH}}$,
C.~Trichard$^{\ref{LLR}}$,
M.~Tsirou$^{\ref{LUPM}}$,
N.~Tsuji$^{\ref{Rikkyo}}$,
R.~Tuffs$^{\ref{MPIK}}$,
Y.~Uchiyama$^{\ref{Rikkyo}}$,
D.J.~van~der~Walt$^{\ref{NWU}}$,
C.~van~Eldik$^{\ref{ECAP}}$,
C.~van~Rensburg$^{\ref{NWU}}$,
B.~van~Soelen$^{\ref{UFS}}$,
G.~Vasileiadis$^{\ref{LUPM}}$,
J.~Veh$^{\ref{ECAP}}$,
C.~Venter$^{\ref{NWU}}$,
P.~Vincent$^{\ref{LPNHE}}$,
J.~Vink$^{\ref{GRAPPA}}$,
H.J.~V\"olk$^{\ref{MPIK}}$,
T.~Vuillaume$^{\ref{LAPP}}$,
Z.~Wadiasingh$^{\ref{NWU}}$,
S.J.~Wagner$^{\ref{LSW}}$,
R.~White$^{\ref{MPIK}}$,
A.~Wierzcholska$^{\ref{IFJPAN},\ref{LSW}}$,
R.~Yang$^{\ref{MPIK}}$,
H.~Yoneda$^{\ref{KAVLI}}$,
M.~Zacharias$^{\ref{NWU}}$,
R.~Zanin$^{\ref{MPIK}}$,
A.A.~Zdziarski$^{\ref{NCAC}}$,
A.~Zech$^{\ref{LUTH}}$,
A.~Ziegler$^{\ref{ECAP}}$,
J.~Zorn$^{\ref{MPIK}}$,
N.~\.Zywucka$^{\ref{NWU}}$,
F.~de Palma$^{\ref{INFNTorino}}$,
M.~Axelsson$^{\ref{OKC},\ref{KTH}}$,
O.J.~Roberts$^{\ref{STI}}$
}

\begin{affiliations}
\small 

\item Centre for Space Research, North-West University, Potchefstroom 2520, South Africa \label{NWU}  
\item Universit\"at Hamburg, Institut f\"ur Experimentalphysik, Luruper Chaussee 149, D 22761 Hamburg, Germany \label{HH}  
\item Max-Planck-Institut f\"ur Kernphysik, P.O. Box 103980, D 69029 Heidelberg, Germany \label{MPIK}  
\item Dublin Institute for Advanced Studies, 31 Fitzwilliam Place, Dublin 2, Ireland \label{DIAS} 
\item High Energy Astrophysics Laboratory, RAU,  123 Hovsep Emin St  Yerevan 0051, Armenia \label{RAU} 
\item Yerevan Physics Institute, 2 Alikhanian Brothers St., 375036 Yerevan, Armenia \label{YPI} 
\item Institut f\"ur Physik, Humboldt-Universit\"at zu Berlin, Newtonstr. 15, D 12489 Berlin, Germany \label{HUB} 
\item University of Namibia, Department of Physics, Private Bag 13301, Windhoek, Namibia, 12010 \label{UNAM} 
\item GRAPPA, Anton Pannekoek Institute for Astronomy, University of Amsterdam,  Science Park 904, 1098 XH Amsterdam, The Netherlands \label{GRAPPA} 
\item Department of Physics and Electrical Engineering, Linnaeus University,  351 95 V\"axj\"o, Sweden \label{Linnaeus} 
\item Institut f\"ur Theoretische Physik, Lehrstuhl IV: Weltraum und Astrophysik, Ruhr-Universit\"at Bochum, D 44780 Bochum, Germany \label{RUB} 
\item Institut f\"ur Astro- und Teilchenphysik, Leopold-Franzens-Universit\"at Innsbruck, A-6020 Innsbruck, Austria \label{LFUI} 
\item School of Physical Sciences, University of Adelaide, Adelaide 5005, Australia \label{Adelaide} 
\item LUTH, Observatoire de Paris, PSL Research University, CNRS, Universit\'e Paris Diderot, 5 Place Jules Janssen, 92190 Meudon, France \label{LUTH} 
\item Sorbonne Universit\'e, Universit\'e Paris Diderot, Sorbonne Paris Cit\'e, CNRS/IN2P3, Laboratoire de Physique Nucl\'eaire et de Hautes Energies, LPNHE, 4 Place Jussieu, F-75252 Paris, France \label{LPNHE} 
\item Laboratoire Univers et Particules de Montpellier, Universit\'e Montpellier, CNRS/IN2P3,  CC 72, Place Eug\`ene Bataillon, F-34095 Montpellier Cedex 5, France \label{LUPM} 
\item IRFU, CEA, Universit\'e Paris-Saclay, F-91191 Gif-sur-Yvette, France \label{IRFU} 
\item Astronomical Observatory, The University of Warsaw, Al. Ujazdowskie 4, 00-478 Warsaw, Poland \label{UWarsaw}
\item Aix Marseille Universit\'e, CNRS/IN2P3, CPPM, Marseille, France \label{CPPM} 
\item Instytut Fizyki J\c{a}drowej PAN, ul. Radzikowskiego 152, 31-342 Krak{\'o}w, Poland \label{IFJPAN} 
\item School of Physics, University of the Witwatersrand, 1 Jan Smuts Avenue, Braamfontein, Johannesburg, 2050 South Africa \label{WITS} 
\item Laboratoire d'Annecy de Physique des Particules, Univ. Grenoble Alpes, Univ. Savoie Mont Blanc, CNRS, LAPP, 74000 Annecy, France \label{LAPP} 
\item Landessternwarte, Universit\"at Heidelberg, K\"onigstuhl, D 69117 Heidelberg, Germany \label{LSW} 
\item Universit\'e Bordeaux, CNRS/IN2P3, Centre d'\'Etudes Nucl\'eaires de Bordeaux Gradignan, 33175 Gradignan, France \label{CENBG} 
\item Oskar Klein Centre, Department of Physics, Stockholm University, Albanova University Center, SE-10691 Stockholm, Sweden \label{OKC} 
\item Institut f\"ur Astronomie und Astrophysik, Universit\"at T\"ubingen, Sand 1, D 72076 T\"ubingen, Germany \label{IAAT} 
\item Laboratoire Leprince-Ringuet, École Polytechnique, UMR 7638, CNRS/IN2P3, Institut Polytechnique de Paris, F-91128 Palaiseau, France \label{LLR} 
\item APC, AstroParticule et Cosmologie, Universit\'{e} Paris Diderot, CNRS/IN2P3, CEA/Irfu, Observatoire de Paris, Sorbonne Paris Cit\'{e}, 10, rue Alice Domon et L\'{e}onie Duquet, 75205 Paris Cedex 13, France \label{APC} 
\item Univ. Grenoble Alpes, CNRS, IPAG, F-38000 Grenoble, France \label{Grenoble} 
\item Department of Physics and Astronomy, The University of Leicester, University Road, Leicester, LE1 7RH, United Kingdom \label{Leicester} 
\item Nicolaus Copernicus Astronomical Center, Polish Academy of Sciences, ul. Bartycka 18, 00-716 Warsaw, Poland \label{NCAC} 
\item Institut f\"ur Physik und Astronomie, Universit\"at Potsdam,  Karl-Liebknecht-Strasse 24/25, D 14476 Potsdam, Germany \label{UP} 
\item Friedrich-Alexander-Universit\"at Erlangen-N\"urnberg, Erlangen Centre for Astroparticle Physics, Erwin-Rommel-Str. 1, D 91058 Erlangen, Germany \label{ECAP} 
\item DESY, D-15738 Zeuthen, Germany \label{DESY} 
\item Obserwatorium Astronomiczne, Uniwersytet Jagiello{\'n}ski, ul. Orla 171, 30-244 Krak{\'o}w, Poland \label{UJK} 
\item Centre for Astronomy, Faculty of Physics, Astronomy and Informatics, Nicolaus Copernicus University,  Grudziadzka 5, 87-100 Torun, Poland \label{NCUT} 
\item Department of Physics, University of the Free State,  PO Box 339, Bloemfontein 9300, South Africa \label{UFS} 
\item Department of Physics, Rikkyo University, 3-34-1 Nishi-Ikebukuro, Toshima-ku, Tokyo 171-8501, Japan \label{Rikkyo} 
\item Kavli Institute for the Physics and Mathematics of the Universe (WPI), The University of Tokyo Institutes for Advanced Study (UTIAS), The University of Tokyo, 5-1-5 Kashiwa-no-Ha, Kashiwa City, Chiba, 277-8583, Japan \label{KAVLI} 
\item Department of Physics, The University of Tokyo, 7-3-1 Hongo, Bunkyo-ku, Tokyo 113-0033, Japan \label{Tokyo} 
\item RIKEN, 2-1 Hirosawa, Wako, Saitama 351-0198, Japan \label{RIKKEN} \and
%% Affiliation of people who left the collaboration
\item Physik Institut, Universit\"at Z\"urich, Winterthurerstrasse 190, CH-8057 Z\"urich, Switzerland \label{MitchellNowAt} 
\item  Institut de Ci\`{e}ncies del Cosmos (ICC UB), Universitat de Barcelona (IEEC-UB), Mart\'{i} Franqu\`es 1, E08028 Barcelona, Spain \label{CerrutiNowAt}
\item Dipartimento Interateneo di Fisica, Politecnico di Bari, Via E. Orabona 4, 70125 Bari, Italy \label{PBari}
\item Istituto Nazionale di Fisica Nucleare - Sezione di Bari, Via E. Orabona 4, 70125 Bari, Italy \label{INFNBari}
\item  Istituto Nazionale di Fisica Nucleare - Sezione di Torino, Via P. Giuria 1, 10125 Torino, Italy \label{INFNTorino}
\item Department of Physics, KTH Royal Institute of Technology, AlbaNova, SE-106 91 Stockholm, Sweden \label{KTH}
\item Science and Technology Institute, Universities Space Research Association, 320 Sparkman Drive, Huntsville, AL 35805, USA \label{STI}

\end{affiliations}

\begin{abstract}
Gamma-ray bursts (GRBs) are brief flashes of gamma rays, considered 
to be the most energetic explosive phenomena in the Universe~\cite{Mezaros2006}. 
The emission from GRBs comprises a short (typically tens of seconds) 
and bright \emph{prompt} emission, followed by a much longer 
\emph{afterglow} phase. During the afterglow phase, the shocked outflow--produced by the interaction between the ejected matter and the circumburst medium--slows down,and a gradual decrease in brightness is observed~\cite{2004IJMPA..19.2385Z}. GRBs typically emit most of their energy via $\gamma$-rays with energies in the kiloelectronvolt-to-megaelectronvolt
range, but a few photons with energies of tens of gigaelectronvolts have been detected by space-based instruments~\cite{GRB130427A_LAT}. However, the origins of such high-energy (above one gigaelectronvolt) photons and the presence of very-high-energy (more than 100 gigaelectronvolts) emission have remained elussive~\cite{PIRON2016617}. Here we report observations of very-high-energy emission in the bright GRB 180720B deep in the GRB afterglow—ten hours after the end of the prompt emission phase, when the X-ray flux had already decayed by four orders of magnitude. Two possible explanations exist for the observed radiation: inverse Compton emission and synchrotron emission of ultrarelativistic electrons. Our observations show that the energy fluxes in the X-ray and γ-ray range and their photon indices remain comparable to each other throughout the afterglow. This discovery places distinct constraints on the GRB environment for both emission mechanisms, with the inverse Compton explanation alleviating the particle energy requirements for the emission observed at late times. The late timing of this detection has consequences for the future observations of GRBs at the highest energies.

\end{abstract}

On July 20th 2018, GRB~180720B triggered the \Fermi Gamma-ray Burst 
Monitor (GBM) at 14:21:39.65\,UT~\cite{GCN22981} ($T_0$) and the 
\Swift Burst Alert Telescope (BAT) 5\,s later~\cite{GCN22973}.  
Multi-wavelength follow-up observations were performed up to 
$T_0+3\times10^5$\,s, notably by the ESO Very Large Telescope (VLT)  
which measured a redshift $z = 0.653$~\cite{GCN22996}. 
In the high-energy (HE) gamma-ray band (100\,MeV--100\,GeV) 
this GRB was also detected by the \Fermi Large Area Telescope (LAT) 
between $T_0$ and $T_0+700$\,s with a maximum photon energy of 5\,GeV 
at $T_0+142.4$\,s~\cite{GCN22980}. No further HE emission was detected in the successive observation windows after $700$\,s. 
The prompt emission phase of GRB~180720B is extremely bright, 
ranking seventh in brightness among the over $\sim 2650$ GRBs 
detected by the \Fermi-GBM up to date (see Methods). With a $T_{90}$ 
(the time in which 90\% of the flux is detected) of $48.9\,\pm\,0.4$\,s, 
it is categorised as a long GRB, typically associated with the 
death of massive stars~\cite{Levan2016}, with an isotropic 
energy release of $E^{\rm iso} = (6.0\,\pm\,0.1)\times10^{53}$~erg 
(50--300\,keV). Observations of this GRB took place using the \Swift 
X-ray Telescope (XRT), identifying a bright afterglow which 
remained detectable until almost 30 days after $T_0$~\cite{SwiftObservations, SwiftObservationsLC} (Fig.~1). 
In terms of energy flux of the X-ray afterglow 
(0.3--10 keV, at $T_0+11$\,h), this GRB ranks second after 
the exceptional GRB\,130427A~\cite{GRB130427A_LAT}.

Observations with the H.E.S.S. array began at $T_0+10.1$\,h and lasted for two hours. The data were analysed using methods optimised for the detection of the lowest energy events revealing a new gamma-ray source 
(Fig.~2.a) with an excess of 119 gamma-ray events and a statistical significance of 5.3\,$\sigma$ (5.0\,$\sigma$ post-trials) (see Methods). The gamma-ray excess is well fit by a point-like source model centred at right ascension  00$\rm ^h$02$\rm ^{min}$07.6$\rm ^s$ and declination $-$02$^\circ$56$'$06$''$ (J2000) with a statistical uncertainty of 1.31$'$, consistent with the measurements at other wavelengths~\cite{GCN23020,GCN22980,GCN22973}.
In order to rule out the association of this source with an unknown steady gamma-ray emitter, such as an active galactic nucleus, or persistent systematic effects, the GRB region was re-observed under similar conditions 18~days after these observations. In total 6.75\,h of data were analysed, resulting in a sky-map consistent with background events (Fig.~2.b).

The flux spectrum detected by H.E.S.S. (100--440\,GeV) was fit with a function of the form $F_{\rm obs}(E) = F_{\rm int}(E)\times e^{-\tau(E,z)}$, with the exponential term accounting for the absorption of photons by the extragalactic background light~\cite{Franceschini2008} and $F_{\rm int}(E) = F_{0,\rm int} (E/E_{0,\rm int})^{-\photonindex_{\rm int}}$ a power-law describing the intrinsic source emission.  The analysis resulted in a photon index $\photonindex_{\rm int} = 1.6\,\pm\,1.2\textrm{(stat.)}\,\pm{0.4}\textrm{(syst.)}$ and a flux normalisation, evaluated at an energy $E_{0,\rm int} = 0.154$\,TeV, of $F_{0,{\rm int}} = (7.52\,\pm\,2.03\textrm{(stat.)}\,^{+4.53}_{-3.84}\textrm{(syst.)})\times 10^{-10}$~TeV$^{-1}$\,cm$^{-2}$\,s$^{-1}$ (see Methods).

The very-high-energy (VHE) flux together with measurements at other wavelengths is shown in Fig.1. Apart from the exceptional flux level, the light curves show a typical power-law behaviour in the X-ray and optical afterglow with a temporal flux decay of the form $F(t)\propto t^{-\alpha}$ with $\alpha_{\rm XRT}=1.29\,\pm\,0.01$ and  $\alpha_{\rm optical} = 1.24\,\pm\,0.02$.  The spectrum measured by the \Fermi-LAT (100~MeV--10~GeV) from $T_0+55$\,s to $T_0+700$\,s is well fit by a power-law model with photon index $\photonindex_{\rm LAT}=2.10\,\pm\,0.10$. The light curve in the same time window is fit by a power-law with a temporal decay index $\alpha_{\rm LAT} = 1.83\,\pm\,0.25$. It is worth noting that $\alpha_{\rm LAT}$ is at $\sim 1 \sigma$  from the mean value of the distribution of the decay indices of long GRBs detected by the \Fermi-LAT~\cite{FermiGRBCat} ($\bar{\alpha}\,_{\rm LAT} = 0.99\,\pm\,0.04$, $\sigma_{\bar{\alpha}} = 0.80\,\pm\,0.07$) and such deviation could largely depend on the time range where $\alpha_{\rm LAT}$ is fitted, potentially in agreement with $\alpha_{\rm XRT}$ and $\alpha_{\rm optical}$.

The detection of VHE gamma-ray emission indicates the presence 
of very energetic particles in the GRB afterglow. This discovery is consistent with efficient gamma-ray emission seen in other astrophysical sources with relativistic plasma outflow, e.g. pulsar wind nebulae or jets emerging from the nuclei of active galaxies. In the case of a GRB afterglow, the particle acceleration likely occurs at the forward shock~\cite{Chevalier_2000} (the compression shock wave propagating through the circumburst material) which should be capable of efficient electron and proton acceleration. As proton radiation processes are characterised by long energy-loss timescales relative to the dynamical timescale, the detected gamma-ray emission is more likely produced by accelerated electrons (see Methods). Therefore, two radiation processes are the most plausible dominant contributions to the VHE spectrum: 
synchrotron emission of an electron population in the local magnetic field~\cite{2009MNRAS.400L..75K} and synchrotron self-Compton (SSC) scattering~\cite{2001ApJ...548..787S,2001ApJ...559..110Z}. In the latter case, the synchrotron photons, thought to dominate the target radiation~\cite{Warren:2017bdm}, are inverse-Compton scattered to higher energies by the same electron population.

The SSC and synchrotron emission-origin scenarios~\cite{2004IJMPA..19.2385Z} place distinctly different demands on the source acceleration efficiency. While an SSC origin requires only multi GeV energy electrons, a synchrotron origin requires an extreme accelerator potentially accelerating beyond PeV energies~\cite{1983MNRAS.205..593G} (see Methods). Furthermore, for GRBs to operate as 10$^{20}$~eV cosmic-rays sources, they must achieve extreme acceleration~\cite{2002PhRvD..66b3005A}. One key distinguishing characteristic between these two emission origins is that SSC predicts the presence of two bumps in the spectral-energy distribution, whose ratio depends on the energy densities in both the electrons and magnetic field, while a synchrotron model implies only a broad single component. A second difference between these processes is the maximum photon energy achievable.

Considering a synchrotron origin of the broadband afterglow energy spectrum, the highest energy for synchrotron emission from electrons in a maximally efficient accelerator is~\cite{2000NewA....5..377A}
$E_{\rm sync}^{\rm max}= 9 \Gamma m c^{2}/ 4\alpha_{\rm F}\approx 100 \Gamma~{\rm MeV}$
(with $\alpha_{\rm F}$ the fine structure constant and $\Gamma$ the bulk Lorentz factor of the forward shock). Thus, for electron-synchrotron emission to 
reach energies beyond $100$~GeV, $10$~hrs after the prompt emission, a late-time $\Gamma$ in excess of 1000 appears required.  Such a scenario is difficult to realise, with robust expectations suggesting a value $\Gamma \sim 20$ at $10$\,h (see Methods).
Alternatively, circumvention of this synchrotron maximum energy limit is possible
for scenarios in which either the coherence length of the magnetic turbulence
is very small, or different magnetic field strengths are present in the acceleration 
and emission zones, or some non-ideal process is responsible for the particle acceleration (see Methods). Regardless of this challenge, this could explain the similarity in the photon index and level of energy flux of the emission seen both at early times by \Fermi-LAT and \Swift-XRT and at late times by H.E.S.S. and \Swift-XRT (Fig.~1). However, the strong requirements for synchrotron emission to extend up to the VHE regime disfavours such origin and the potential onset of a new SSC component should be considered.

The SSC scenario has the advantage that the emission up to VHE at late times is energetically much more easily 
achievable~\cite{Warren:2018lyx}, leading 
to the expectation of a new spectral component at VHE. The H.E.S.S. spectral fit 
constraints (Extended Data Fig.~1) are consistent with such a 
possibility within present uncertainties. Despite this advantage, the potential onset 
of inverse Compton emission within the Klein-Nishina (KN) regime faces challenges 
(see Methods). Specifically, 
beyond the gamma-ray energy where this sets in, a softer spectral slope and a 
different brightness evolution of this component~\cite{2009ApJ...703..675N} are expected. 
However, interestingly, the presence of synchrotron emission with a hard photon
index extending below keV energies can delay the onset of the full KN transition
sufficiently to higher energies (see Methods), beyond that of the very-high energy detection. The detection of this hard extended synchrotron emission component thus
delivers additional supporting evidence for an SSC origin.

This VHE discovery undoubtedly opens a key channel to the understanding of the GRBs afterglow phenomena. This measurement proves complementary to the VHE-afterglow emission detected in GRB~190829A~\cite{HESSatel} and the prompt-to-early afterglow emission measured in GRB~190114C by the MAGIC telescopes~\cite{MAGICGRB}, providing insight into the nature of GRBs and their VHE detectability. We estimate that future instruments, such as the Cherenkov Telescope Array~\cite{CTAbook}, will allow up to 3 more GRB afterglow detections per year in the VHE domain than previously anticipated (see Methods), significantly improving our understanding of GRBs over a diverse range of timescales.

\newpage 

{\bf References}
\vspace{1em}

%\bibliographystyle{naturemag}
%\bibliography{references}

% ------------------------------------------------------------------------------

% ------------------------------------------------------------------------------

\newpage

% ------------------------------------------------------------------------------
% MWL Light curve figure 
\begin{figure}[ht!]
  \includegraphics[width=1.0\textwidth]{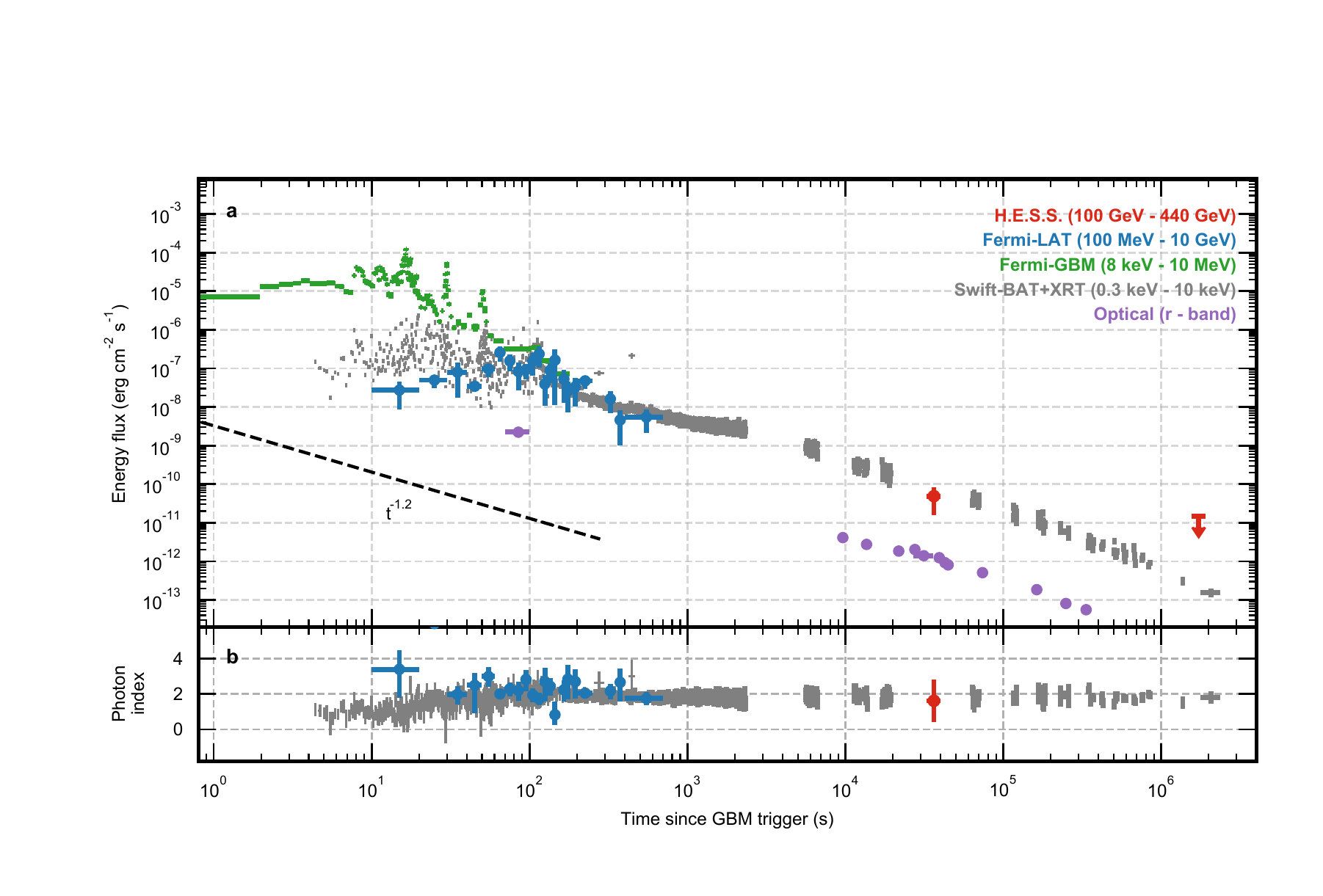}
  \caption{{\bf Multi-wavelength light curve of GRB~180720B.} 
  Panel \textbf{a} shows the energy flux light curve detected by the \Fermi-GBM (Band fit, green), \Fermi-LAT (power-law, blue), H.E.S.S energy flux and 95\% Cl L.(power-law intrinsic, red), optical r-band (purple). 
  The \Swift-BAT spectra (15 keV--150 keV) extrapolated to the XRT band (0.3--10 keV) 
  for a combined light curve (grey). 
  The black dashed line indicates a temporal decay with $\alpha = -1.2$. Panel \textbf{b} shows the photon 
  index of the \Fermi-LAT, \Swift and 
  H.E.S.S. spectra. Error bars correspond to 1~$\sigma$ standard deviation.
  }
  \label{fig:lightcurve}
\end{figure}

% ------------------------------------------------------------------------------

% ------------------------------------------------------------------------------
% Gamma-ray image of GRB~180720B

\begin{figure}[ht!]
  \includegraphics[width=1.0\textwidth]{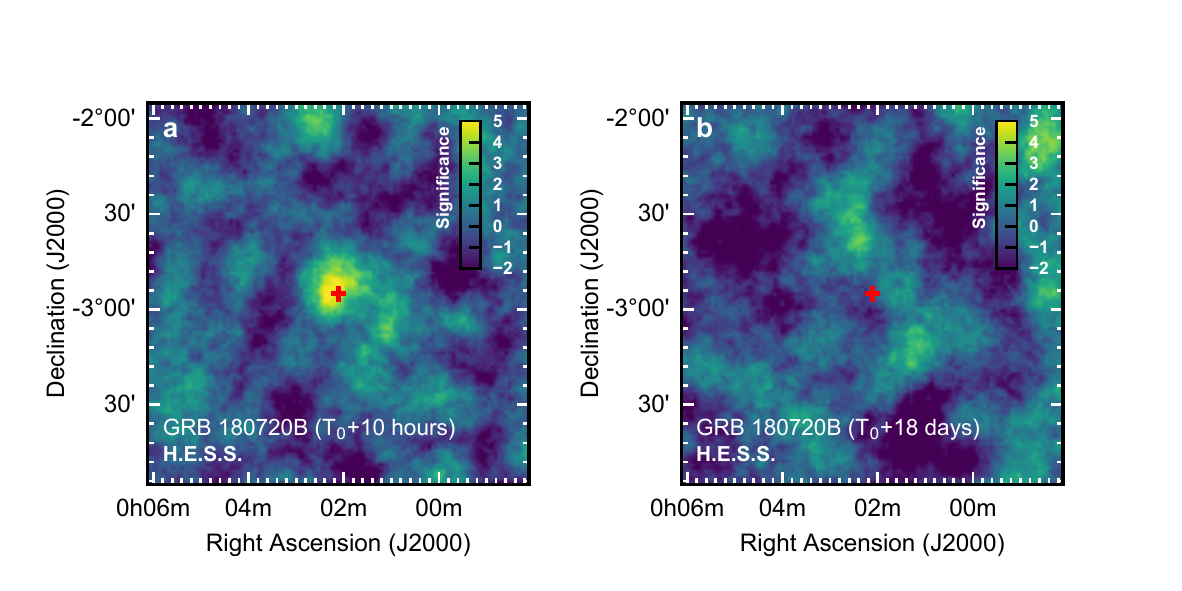}
  \caption{{\bf Very-high-energy gamma-ray image of GRB~180720B.} 
  Significance map of the GRB~180720B field as observed by H.E.S.S. 
  Panel \textbf{a} corresponds to the observation made at $T_0+10.1$\,h 
  for a total of 2\,h. Panel \textbf{b} 
  shows the same region of the sky as observed during consecutive 
  nights between $T_0+18.4$\,days and $T_0+24.4$\,days. 
  The red cross indicates the position reported by the optical 
  telescope ISON-Castelgrande~\cite{GCN23020}.
  }
  \label{fig:skymap}
\end{figure}

% ------------------------------------------------------------------------------

\newpage

\begin{center}
    \textbf{Methods}
\end{center}

\renewcommand\tablename{Extended Data Table}
\renewcommand\figurename{Extended Data Figure}
\setcounter{figure}{0}

\subsection{H.E.S.S. and the GRB follow-up programme}
The observations presented in this paper were performed using the H.E.S.S. array of imaging atmospheric Cherenkov telescopes situated at an altitude of 1800\,m in the Khomas highlands of Namibia. H.E.S.S. is sensitive to gamma-rays in the energy range from tens of GeV to tens of TeV. It consists of five Cherenkov telescopes, four of 108\,m$^{2}$ in mirror area placed on a square configuration of side length 120\,m (CTs 1--4) and a single telescope at the centre (CT5) with a mirror area of  614\,m$^{2}$. Thanks to its low energy threshold and fast slewing (200 degrees per minute~\cite{drivesystem}), CT5 is well suited for the observation of soft-spectrum transient sources.

H.E.S.S. maintains an active transient source observation programme, of which gamma-ray bursts (GRB) are an important component. To ensure a fast reaction to GRB alerts, H.E.S.S. is connected to the gamma-ray coordinates network (GCN)~\cite{GCNsystem}, which rapidly distributes alerts and observational information from space and ground-based facilities. The target of opportunity observation system in H.E.S.S. performs the selection, filtering and processing of these alerts based on source observability and significance, aiming to trigger on bright, precisely located, and nearby bursts. Alerts are followed-up in two different observation modes. Observations are triggered in \emph{prompt} mode when the GRB position is observable from the H.E.S.S. site at the time that the alert is received.  In this case, the current observation schedule is interrupted and the array is automatically re-pointed to the GRB location. On the other hand, \emph{afterglow} observations take place for GRBs which only become observable at a later time and are scheduled manually and triggered by a burst advocate. This was the case for GRB~180720B, observed from $T_0+10.1$~h when the burst position rose above 45$^\circ$ in elevation (below this elevation GRBs are typically not observed due to the rapid increase in energy threshold). Re-observations were carried out at $T_0$+18\,days, after the end of an intervening moonlight period.

\subsection{H.E.S.S. data analysis}
\label{hessanalysis}
To reach the lowest possible energies in the analysis presented here, we use only data from the single large telescope (CT5). However, this energy threshold reduction comes at the cost of some angular resolution and sensitivity loss~\cite{HESSIIcrab}. In total we present here two hours of observations taken in \emph{wobble mode}~\cite{bgmodeling} with the telescope pointing direction at an offset of 0.5$^{\circ}$ from the position provided by \Swift-BAT~\cite{GCN22973}.  This observation was made at a mean zenith angle of 31.5$^{\circ}$ for a total live-time of 1.8 h. 

In order to ensure that a potential GRB signal is not diminished by an excessive number of statistical trials, the data analysis is subject to a strict \emph{unblinding} procedure. The first step in this unblinding is an inspection of the low-level data, as some calibration artefacts can directly lead to the creation of spurious sources in the field of view. 
Checks are made on the fractional event participation of each camera pixel (to ensure single faulty pixels do not dominate the events), the pixel pedestal values, and the distribution of events within the field of view. 
Once these checks are completed, with no artefacts found, the event properties are reconstructed using the \emph{ImPACT}~\cite{ImPACT, ImPACTmono} maximum likelihood-based fitting technique. Background cosmic-ray events are rejected using a neural network based scheme~\cite{HESSIImonoHillas}.
The residual background contamination level of the source region (\emph{ON} and \emph{OFF} events) and the ratio of the on-source time to the off-source time ($\alpha_{\rm Exp.}$) is then estimated using the \emph{ring} method for the production of maps and \emph{reflected-region} method when performing the spectral extraction~\cite{bgmodeling}. Full analysis and checks are performed using an additional independent calibration and data analysis chain~\cite{modelanalysis}, serving as a cross-check of all the results presented. 
 
 The source significance is computed by using a maximum-likelihood ratio test based on the number of events coming from the source (ON) and the background (OFF) for a given on-source to off-source time ($\alpha$)~\cite{1983ApJ...272..317L}.  For the \emph{ring} method, the number of ON and OFF events is 544 and 4740, and $\alpha_{\rm Exp.} = 0.09$ resulting in a significance of detection of 5.3~$\sigma$. Similarly the \emph{reflected-region} measured 544 and 3998 events, and $\alpha_{\rm Exp.} = 0.11$ for a significance of 4.6~$\sigma$ which is verified by the cross-check analysis providing 651 and 5200 of ON and OFF events and $\alpha_{\rm Exp.} = 0.10$  with a significance of 4.5~$\sigma$.

% Need to provide stats here now, also for cross check? - RDP

The source morphology is fit with a 2D-likelihood procedure by assuming point-like and Gaussian source models convolved with the expected energy-dependent point spread function (obtained from simulations) and the measured source spectrum. Both source models proved to be compatible with the morphology of the discovered source, with no significant preference for source extension shown.

Spectral analysis is performed using the forward-folding method~\cite{spectra},  which corrects for the limited energy resolution of the single-telescope event reconstruction. The measured source spectrum is obtained by fitting a simple power-law model of the form  $F_{\rm obs}(E) = F_{0,\rm obs} (E/E_{0,\rm obs})^{-\photonindex_{\rm obs}}$, where $F_{0,\rm obs}$ is the flux normalisation, $\photonindex_{\rm obs}$ the photon index, and $E_{0,\rm obs}$ the reference energy. However, due to the absorption of the most energetic photons by the extragalactic background light (EBL), the apparent photon index of this source will be somewhat steeper than the intrinsic behaviour. The intrinsic spectrum $F_{\rm int}(E)$ is therefore obtained by fitting the measured spectrum to an attenuated power-law model $F_{\rm obs}(E) = F_{\rm int}(E)\times e^{-\tau(E,z)} = F_{0,\rm int} (E/E_{0, \rm Int})^{-\photonindex_{\rm int}} \times e^{-\tau(E,z)}$, where the last term in the equation corresponds to the EBL absorption coeficient predicted~\cite{Franceschini2008} for a redshift of 0.653. The best-fit spectra together with the spectral points are shown in Extended Data Fig.~1, and the spectral parameters are summarised in Extended Data Table~1.

Systematic uncertainties in the fitted spectra are determined accounting for a 15\% uncertainty in the reconstructed energy due to possible variations in the measured Cherenkov light yield~\cite{2017A&A...600A..89H}. The measured energy is systematically shifted by $\pm$15\% and the whole spectral fitting procedure is redone. In addition to this, short dips in the trigger rate, at the level of 30\%, were identified in the data, most likely attributed to the presence of high altitude clouds. In order to assess the effect of these, the time windows containing such trigger rate features (21.7\,min total) were removed from the data and the standard analysis described above was performed on the reduced data set. From this, we conclude an additional systematic underestimation of 32\% and 4.8\% in the measured normalisation and photon index respectively.  These two sources of systematic uncertainty are considered to be independent and therefore added in quadrature for a total systematic uncertainty estimation. 

The intrinsic spectrum was obtained with a chosen EBL model~\cite{Franceschini2008}. To determine the level by which this choice influences the results presented in this work, the data were re-analysed using three additional EBL models~\cite{Dominguez2011, Finke2010, Gilmore2012}, each one employing a different approach to predict the overall EBL level~\cite{Dominguez2011}.
The absorption coefficient for a redshift of 0.653 within the energy range of the detected emission does not present significant deviations between these models considered(Extended Data Fig.~2 extended data). When employing these EBL models for the spectral fit, a change in the reported normalisation and index of maximum 55.3\% and 27\% respectively was found. The statistical uncertainty on the fitted spectra remains as the biggest source of uncertainty in the results.

\subsection{Trial correction}% and false alarm rate.} 

Since 2012, H.E.S.S. has performed five additional follow-up observations of well localised GRBs (\Swift and \Fermi-LAT alerts) using only CT5 (similar to the observations presented here). The significance distribution of this sample (excluding GRB~180720B) is consistent with pure statistical fluctuations. Therefore, a post-trial significance for GRB~180720B can be assessed by accounting for these GRBs previously observed. This results in a post-trial significance of 4.3~$\sigma$ (\emph{reflected-region} method) and 5.0~$\sigma$ (\emph{ring} method). 
As the analysis of GRB~180720B was performed once under the aforementioned unblinding procedure, no additional trials have been added to the results presented here.

\subsection{Background systematic effects}

Systematic effects on the sky map background (Fig.~2) were determined by measuring the significance distribution when excluding the source region. While a normal distribution is expected, a width of 1.09 was measured in this significance distribution, therefore adding a slight shift to the reported significance of the \emph{ring} method (used in the production of sky maps). The corrected significance when accounting for such effects is 4.9~$\sigma$ (4.7~$\sigma$ post-trial). Nonetheless, this measured distribution highly depends on the parameters of the \emph{ring} method and should be subject to statistical uncertainties.

\subsection{\Fermi data analysis} 
The \Fermi-GBM data for GRBs are publicly available  through the GBM Burst Catalog at HEASARC~\cite{FermiGBRST}. For GRB~180720B the available time-tagged events (TTE) of those detectors having the best viewing angle to the \Swift-XRT position, namely n6, n7, nb, and b1, were analysed. Temporally-resolved energy flux data points (Fig.~1) were obtained with the {\it RMfit} analysis software~\cite{FermiGBMSoftware} by combining TTE data from all four detectors into 256~ms bins in the energy range from 8 keV to 10 MeV.

The \Fermi-LAT data analysis was performed using the {\it ``Pass8''}~\cite{Atwood:2013rka} processed events. We used the P8R3\_TRANSIENT010E event class that is suitable for transient source analysis, and the corresponding instrument response functions\cite{FermiLATPerformance}. Events were selected from $T_0$ to $T_0+700$\,s in the standard GRB-analysis energy range from 100 MeV to 100 GeV over a region of 10$^{\circ}$ around the \Swift-XRT localisation. Event selection, quality cuts and data analysis were performed with the standard \emph{FermiTools}~\cite{FermiTools} software. The source detection over the full duration was determined by a likelihood analysis providing a test-statistic value of TS~$\sim 600$, which corresponds to a significance $\sigma \sim 25$ ($\sigma\sim\sqrt{TS})$.

Since the highest energy photon detected has an energy of 5 GeV (at $T_0+142.4$\,s) the temporally-resolved energy flux data points (Fig.~1) were computed in the energy range from 100 MeV to 10 GeV. The analysis model included the Galactic interstellar emission model (gll\_iem\_v06.fits) and the relative isotropic diffuse emission templates provided by the \Fermi-LAT collaboration~\cite{Acero2016}, 
where the normalisation of the latter was left free to vary. The spectrum for each bin was fit to a single power-law model  $F(E) = F_0 \times \left( \frac{\rm E}{\rm E_0} \right)^{-\photonindex}$ having the flux normalisation $F_0$ and the photon index $\photonindex$ as free parameters. As no emission with energies $>10$\,GeV is detected, no additional term in the spectra accounting for EBL absorption~\cite{Franceschini2008}  was required. 
The temporal decay $\alpha_{\rm LAT}$ was fit to a power-law model using a least-squares technique applied from $T_0+55$\,s to $T_0+700$\,s, in order to assure no contamination of the \emph{prompt} emission as observed by the \Fermi-GBM and \Swift-BAT, obtaining a reduced $\chi^2 = 0.63$ (d.o.f.$=14$).

\subsection{Optical data}

The optical data shown in Fig.~1 were compiled from the GCN circulars of observations performed in the r-band by the following instruments:
Kanata~\cite{GCN22977}, MITSuME~\cite{GCN22983}, TSHAO~\cite{GCN22979}, MASTER-K~\cite{GCN23023} , MASTER-I, ISON-Castelgrande~\cite{GCN23020}, OSN~\cite{GCN22985}, LCO~\cite{GCN22976}, and KAIT~\cite{GCN23033}. The reported temporal decay index $\alpha_{\rm optical}$ is measured from $T_0+9642$~s to $T_0+3.35\times 10^{5}$~s by performing a power-law fit with a $\chi^2$ fitting procedure.

\subsection{\Swift data}

The \Swift data are publicly available through the \Swift online repository~\cite{SwiftOnline}. The temporally resolved energy-flux data shown in Fig.~1 were obtained using the \emph{Burst Analyser} tool~\cite{SwiftObservations,SwiftObservationsLC}. Data are rebinned for a signal-to-noise ratio (SNR) of 7 and systematic uncertainties are included. The temporal decay reported here ($\alpha_{\rm XRT}$) is obtained from $T_0+2200$~s to $T_0+3.05\times$10${^5}$~s and corresponds to the fourth break in the light curve as identified from the fitting procedures of the \Swift-XRT tools. 

\subsection{CTA detectability prospects} 

Considering CTA to be an order of magnitude more sensitive than the H.E.S.S. array implies the ability to detect energy fluxes $\sim 10$ times fainter than that of GRB~180720B at VHE. If the VHE energy flux equals that detected by \Swift-XRT, as suggested by our measurements (Fig.~1), we estimate an occurrence of 3 GRBs per year above such assumed flux and therefore detectable by CTA (Extended Data Fig.~3). Such number could be increased for follow-up observations at earlier times. By assuming a temporal decay value $\alpha=1.2$ ($F(t)\propto t^{-\alpha}$) for all the GRB afterglows detected by \Swift-XRT~\cite{Nousek:2005fm} an extrapolation of the 11 hours energy flux to the one expected at 5 hours provides a detectability prospect of $\sim 10$ GRBs per year at such follow-up delay times. It should be noted however, that the presence of VHE emission could also be dependent on the GRB environment~\cite{Vurm:2016qqi} and such influence is not considered in this estimation.

\subsection{The Bulk Lorentz Factor} depends on two factors: the released energy and the density of the circumburst
medium~\cite{1976PhFl...19.1130B}, \(\Gamma = \sqrt{{E^{\rm iso}}/{Mc^2}}\),
where \(E^{\rm iso}\) is the equivalent isotropic energy and \(M\) is the total mass swept-up by the shock.
The latter depends on the nature of the circumburst environment: \(M=(4\pi/3) R^3 n m_p \) for a homogeneous medium (here \(n\) is the medium number density and \(R\) is the shock radius) or \(M=\dot{M}_* R/v_*\) for a shock propagating in a constant velocity wind (here \(\dot{M}_*\) and \(v_*\) are the wind mass-loss rate and velocity, respectively). The shock radius depends on the detection time as \(R\simeq A_{\rm R}\Gamma^2 tc / (1+z)\), where \(A_{\rm R}=8\) and \(4\) for homogeneous medium and wind environments, respectively. Thus, for GRB 180720B (\(t=10\rm\,h\), \(z=0.653\), and \(E^{\rm iso}\simeq10^{54}\rm\,erg\)) one obtains \(\Gamma=15n_0^{-1/8}\) (here \(n_0=n/{\rm cm}^{-3}\)) for a homogeneous medium
or  \(\Gamma=20 \dot{M}_{*,-5}^{-1/4}v_{*,3.3}^{1/4}\) (here \(\dot{M}_{*,-5}=\dot{M}_*/(10^{-5}M_{\rm sun}/\rm yr)\) and \(v_{*,3.3}=v_* / ( 2000\rm\, km/s)\)) for a wind environment.

\subsection{Non-thermal process efficiency}, \(\kappa=t_{\rm dyn}/t_{\rm cool}\), depends on the ratio of the shock
dynamic time-scale, \(t_{\rm dyn}= R /(c\Gamma)\), to the cooling time, \(t_{\rm cool}\), which depends on radiation
mechanism, density of the target, and the energy of the parent particles. For hadronic
processes~\cite{2004vhec.book.....A}, which include proton-proton (\(pp\)) and photon-meson (\(p\gamma\)) channels, the
radiation efficiency is \(\kappa_{pp}\sim10^{-7}\left({R/10^{18}\rm\,cm}\right)n_0\) and
\(\kappa_{
  p\gamma}\sim3\times10^{-4}\left({\Gamma/20}\right)^{2}\left({R/10^{18}\rm\,cm}\right)\left(\kappa/10^{-2}\right)\left({E_{\rm
      t}/1\rm\,keV}\right)^{-1}n_0\) (here $\kappa$ is the total radiative efficiency, and $E_{\rm t}$ is the
peak frequency of the soft emission component). These low efficiencies favor the electromagnetic
processes~\cite{1970RvMP...42..237B}. The efficiency of the synchrotron channel for the emission detected in the VHE
band, \(E\sim100\rm\,GeV\), is
\(\kappa_{\rm sync}=5\times10^{7}\left({m_e/m}\right)^{5/2}\left({R/10^{18}\rm\,cm}\right)\) \(\left({\eta_{\rm
      B}/0.1}\right)^{3/4}\eta_{\rm turb}^{-1/2}n_0^{3/4}\), where \(m_e\) and \(m\) are electron and emitting particle
masses; \(\eta_{\rm B}\) is a fraction of the internal energy contained in the magnetic field; and
\(\eta_{\rm turb}=\max\left(1,R_{\rm Lar}^{\rm nr}/\lambda_{\rm cor}\right)\) defines the shift of the peak energy if a
charged particle interacts with turbulent magnetic field~\cite{2013ApJ...774...61K} (here \(R_{\rm Lar}^{\rm nr}\) is the
non-relativistic Larmor radius and \(\lambda_{\rm cor}\) is the magnetic field correlation length).
% The inverse Compton efficiency is determined by the density of the target photons.
If the inverse Compton scattering proceeds in the Thomson regime, then  \(\kappa_{\rm IC}=3\left(\Gamma/20\right)\left(\kappa/10^{-2}\right)\left({R/10^{18}\rm\,cm}\right)\) \(\left({E_{\rm t}/1\rm\,keV}\right)^{-1/2}n_0\). Efficiencies larger than 1 indicate that particle cooling occurs faster than the source dynamical timescale, and is therefore highly efficient.

\subsection{Synchrotron emission} is characterised by the highest radiation efficiency, but this emission component peaks below this limiting energy:
\(E_{\rm sync}^{\rm max} = 100 {\Gamma} \left({m}/{m_e}\right) \eta_{\rm turb}\left({B_{\rm em}}/{B_{\rm ac}}\right)\left({E_{\rm ac}}/{B_{\rm ac}}\right)\rm\,MeV\).
Here \(B_{\rm em}\) and \(B_{\rm ac}\) are magnetic field strengths at the emitter and accelerator sites, respectively. The accelerating electric field, \(E_{\rm ac}\), is smaller than the magnetic field, \(E_{\rm ac}<B_{\rm ac}\), if the particle acceleration proceeds in the ideal magnetohydrodynamic (MHD) flows~\cite{2002PhRvD..66b3005A}. 
Thus, the production of VHE gamma rays via electron synchrotron emission requires a large Lorentz factor, \(\Gamma>10^3\); a very small-scale magnetic turbulence, \(\lambda_{\rm cor}<10^{-2}R_{\rm Lar}^{\rm nr}\); a large change of the magnetic field strength, \(B_{\rm em}>10^2B_{\rm ac}\); particle acceleration to operate in the non-ideal MHD regime;  or a combination of these factors. Proton synchrotron alleviates these requirements in expense of a significantly lower radiation efficiency. Whilst proton synchrotron emission out-competes  other hadronic radiation processes in terms of efficiency~\cite{2001ApJ...559..110Z}, its efficiency is still significantly smaller  than that for electrons. Thus proton synchrotron emission is expected to only give rise to a subdominant emission component within the VHE band.

\subsection{The energy of particles emitting in the VHE regime} depends on the dominant radiation mechanism and ejecta
properties. In the case of a synchrotron origin scenario the particle energy is determined by three important factors: the
shock Lorentz factor, the strength of the magnetic field, and the turbulence scale. While the first factor, \(\Gamma\simeq20\), is relatively well defined by the epoch of the H.E.S.S. observation, the magnetic field strength and possibility of the realisation of small-scale turbulence remains highly uncertain. The internal energy density, \(\sim0.1(\Gamma/20)^2n_0\rm\,J/m^3\), suggests that a Gauss-strength magnetic field is expected for the case of energy equipartition between the magnetic field and particles. We note, however, that significantly smaller plasma magnetisation is discussed in the literature~\cite{2014ApJ...785...29S}, corresponding to several orders of magnitude weaker magnetic field strengths. Assuming
that synchrotron emission beyond the 100~MeV energy limit in co-moving frame can be achieved, the energy of the emitting electrons can be estimated as  \(E_{e}\simeq 4 ({E /100\rm\,keV})^{1/2}(\Gamma/20)^{-1/2}({B/0.1\rm\,G})^{-1/2}\eta_{\rm turb}^{-1/2}\rm\,TeV\). The production of \(100\rm\,GeV\) gamma rays through a synchrotron scenario therefore requires electrons of ultra-high-energy, \(E_{e}\sim4\rm\,PeV\), unless a configuration with a very small-scale turbulence is present. The energy of particles that provide the dominant contribution to the IC emission depends strongly on the spectrum of the target photons and bulk Lorentz factor. An electron with energy \(E_{e}\) up-scatters a target photon with energy $E_{\rm t}$ to energy \(\min(E_{\rm t} (E_{e}/m_{e}c^2)^2,\Gamma E_{e})\). For target photons detected in the X-ray energy band \(E_{\rm t}\sim 1 \rm\,keV\),  electrons with energy \(E_{e}\sim 10 \rm\,GeV\), which in the lab frame have an energy of hundred of GeV, can produce gamma rays that are detected in the VHE energy band. 

\subsection{Target photons } of very different energies can be up-scattered to gamma rays of the same energy. This can be of particular relevance for VHE detected from GRBs, with both target photons and non-thermal electrons likely possessing broad energy distributions. Assuming a power-law distribution for the target photon flux, $ d n/dE_{t}\propto E_{t}^{-\gamma}$, and for electrons, $dn_{\rm e}/dE_{e}\propto E_{e}^{-\gamma_{\rm e}}$ one finds that the relative contribution to the gamma-ray emission depends on the electron energy as \(\propto (1- E/
(E_{e}\Gamma)) E_{e}^{2\gamma-\gamma_{\rm e}}\). For the sake of simplicity, just a single high-energy term  in the cross-section was accounted for (resulting in the factor \(1- E/(E_{e}\Gamma)\)), sufficient for a qualitative study. However, the obtained dependence shows that for a reasonable range of photon and electron indices, \(1.5<\gamma,\gamma_{\rm e}<3\), the highest-available-energy electrons may provide an important contribution to the gamma-ray energy band by up-scattering photons with energies within the infrared (IR) to ultra-violet (UV) range.

\subsection{The Klein-Nishina cutoff} is a significant reduction of the Compton cross-section that occurs when
\(E_e E_{\rm t}\gtrsim \Gamma m_e^2c^4\), where \(E_e\) and \(E_{\rm t}\) are the electron and target
photon energies in the co-moving frame and laboratory systems, respectively. This results in a softening of the
gamma-ray spectrum that occurs for
\(E \gtrsim 50 \left({\Gamma}/{20}\right)^2\left({E_{\rm t}/{1~{\rm keV}}}\right)^{-1}\)~GeV.  Since
  typically GRB synchrotron spectral energy distribution peaks in the keV band, the inverse Compton component detected at late afterglow
  phases may be affected by the Klein-Nishina cutoff, resulting in reduced fluxes and steeper spectra. This may appear to be in contradiction with a relatively hard intrinsic spectral index, \(\gamma_{\rm int}\sim1.6\) inferred from the
  H.E.S.S. measurement. There are, however, two effects that can result in a spectral hardening at energies around the cutoff: (i) the up-scattering
  of low energy IR to UV photons, giving an intrinsic VHE component with the same slope as seen in the hard X-ray band;
  (ii) and hardness of the electron spectrum at GeV energies, where adiabatic losses likely render the electron spectrum
  hard. The search for consistency within this framework of the hard VHE spectrum with the SSC scenario, however, requires detailed dedicated simulations, which are beyond the scope of 
  this observational paper.
%: the inverse Compton component is expected to be affected the Klein-Nishina cutoff resulting in smaller fluxes and steeper spectra.

\newpage

{\bf References}
\vspace{1em}

% ------------------------------------------------------------------------------

% ------------------------------------------------------------------------------

\newpage

\begin{addendum}
 \item The support of the Namibian authorities and of the University of Namibia in facilitating the construction and operation of H.E.S.S. is gratefully acknowledged, as is the support by the German Ministry for Education and Research (BMBF), the Max Planck Society, the German Research Foundation (DFG),  the  Helmholtz Association,  the Alexander  von  Humboldt Foundation,  the  French  Ministry  of  Higher  Education,  Research  and  Innovation,  the  Centre  National  de  la  Recherche  Scientifique  (CNRS/IN2P3  and CNRS/INSU), the Commissariat a l'energie atomique et aux energies alternatives (CEA), the U.K. Science and Technology Facilities Council (STFC), the Knut and Alice Wallenberg Foundation,  the National Science Centre,  Poland grant no. 2016/22/M/ST9/00382, the South African Department of Science and Technology and National Research Foundation, the University of Namibia, the National Commission on Research, Science \& Technology of Namibia (NCRST), the Austrian Federal Ministry of Education, Science and Research and the Austrian Science Fund (FWF), the Australian Research Council (ARC), the Japan Society for the Promotion of Science and by the University of Amsterdam.  We appreciate the excellent work of the technical support staff in Berlin, Zeuthen, Heidelberg, Palaiseau, Paris, Saclay, T\"ubingen and in Namibia in the construction and operation of the equipment.  This work beneted from services provided by the H.E.S.S. Virtual Organisation, supported by the national resource providers of the EGI Federation.
 
 The \textit{Fermi}-LAT Collaboration acknowledges support for LAT development, operation and data analysis from NASA and DOE (United States), CEA/Irfu and IN2P3/CNRS (France), ASI and INFN (Italy), MEXT, KEK, and JAXA (Japan), and the K.A.~Wallenberg Foundation, the Swedish Research Council and the National Space Board (Sweden). Science analysis support in the operations phase from INAF (Italy) and CNES (France) is also gratefully acknowledged. This work performed in part under DOE Contract DE-AC02-76SF00515.
 
\item[Data and Code Availability Statement] 
The H.E.S.S. raw data and the code used in this study are not public but belong to the H.E.S.S. collaboration. All derived higher level data that are shown in plots will be made available on the H.E.S.S. collaboration’s web site upon publication of this study.
 
Data and analysis code from the Fermi-GBM and LAT instruments are publicly available. Links to the data and software are provided in the analysis section.

This work also made use of data supplied by the UK Swift Science Data Centre at the University of Leicester (\texttt{http://www.swift.ac.uk/archive/}).

\item[Author information]

 \item[Author contributions] R.~D.~Parsons, Q.~Piel, E.~L.~Ruiz-Velasco have analysed the H.E.S.S. data, interpreted the data and prepared the manuscript. E. Bissaldi analysed the \Fermi data, interpreted the data and prepared the manuscript. C. Hoischen implemented the response system for the GRBs follow-up of H.E.S.S. We acknowledge the help of A.~M.~Taylor, F.~Aharonian and D.~Khangulyan on interpreting the results and preparing the manuscript. The  whole  H.E.S.S.  collaboration  has  contributed  to  the  publication  with  involvement at various stages ranging from the design, construction and operation of the instrument, to the development and maintenance of all software for data handling,  data  reduction  and  data  analysis.   All  authors  have  reviewed,  discussed, and commented on the present results and on the manuscript.

 \item[Competing Interests] The authors declare that they have no
competing financial interests.

\item[Correspondance and request for materials] should be addressed to E.R.-V. (contact.hess@hess-experiment.eu).

\end{addendum}

\begin{figure}[H]
    \centering
    \includegraphics[width = 1.0\textwidth]{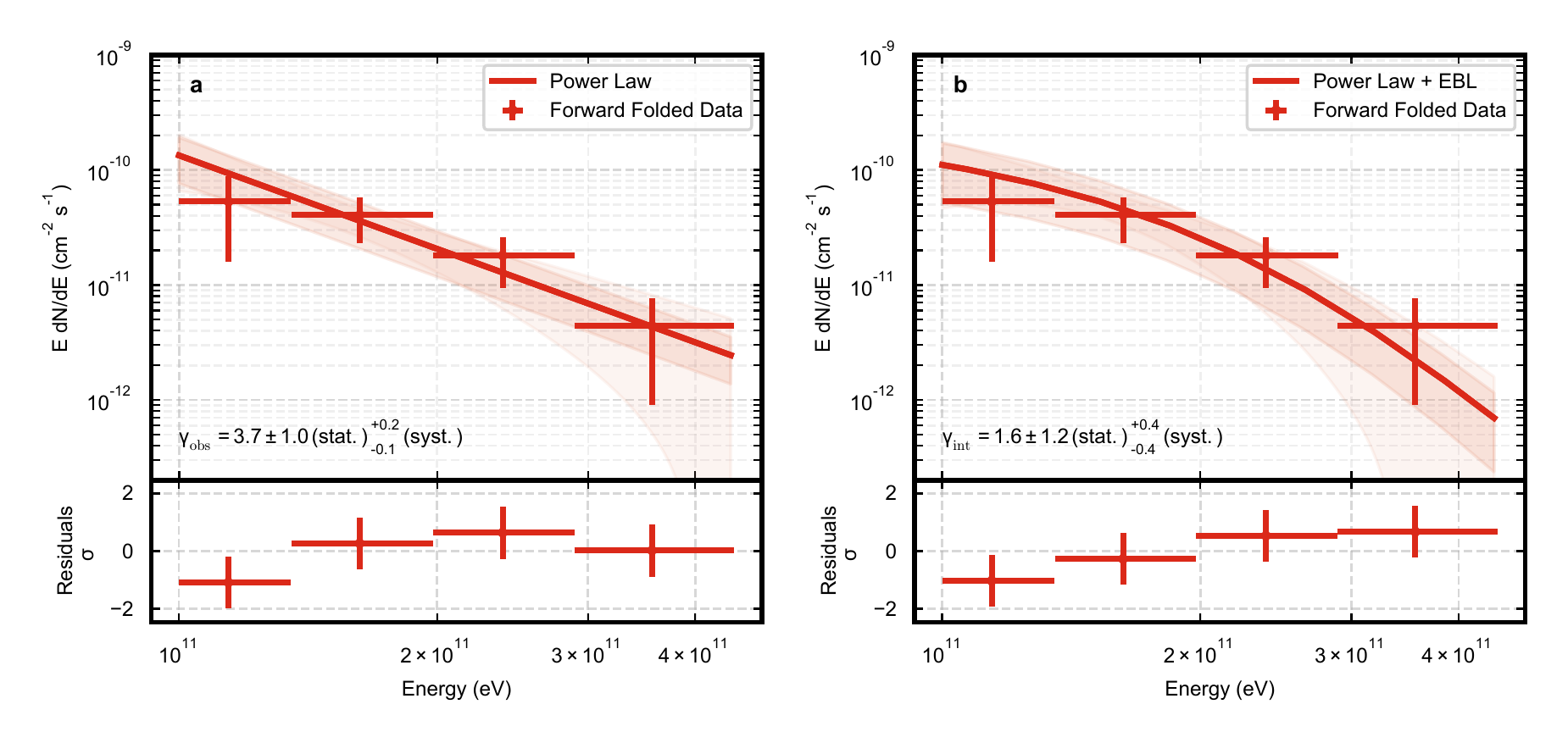}
    \caption{\textbf{VHE spectral fit of GRB~180720B.} H.E.S.S. spectral fit to the measured emission in the energy range of 100--440 GeV. Panel \textbf{a}: Fit assuming a simple power-law model (with photon index $\photonindex_{\rm obs}$). Panel \textbf{b}: Fit assuming a power-law model (with photon index $\photonindex_{\rm int}$) with EBL attenuation for a source at $z=0.653$~\cite{Franceschini2008}. In both cases the residual data points with 1~$\sigma$ uncertainties are obtained from the forward-folded method. The shaded areas show the statistical and systematic uncertainties in each fit ($1~\sigma$ confidence level). The bottom panels show the significance of the residuals between the fitted model and the data points.}
    \label{fig:spectrum}
\end{figure}

\begin{figure}[H]
    \centering
    \includegraphics[width = 0.5\textwidth]{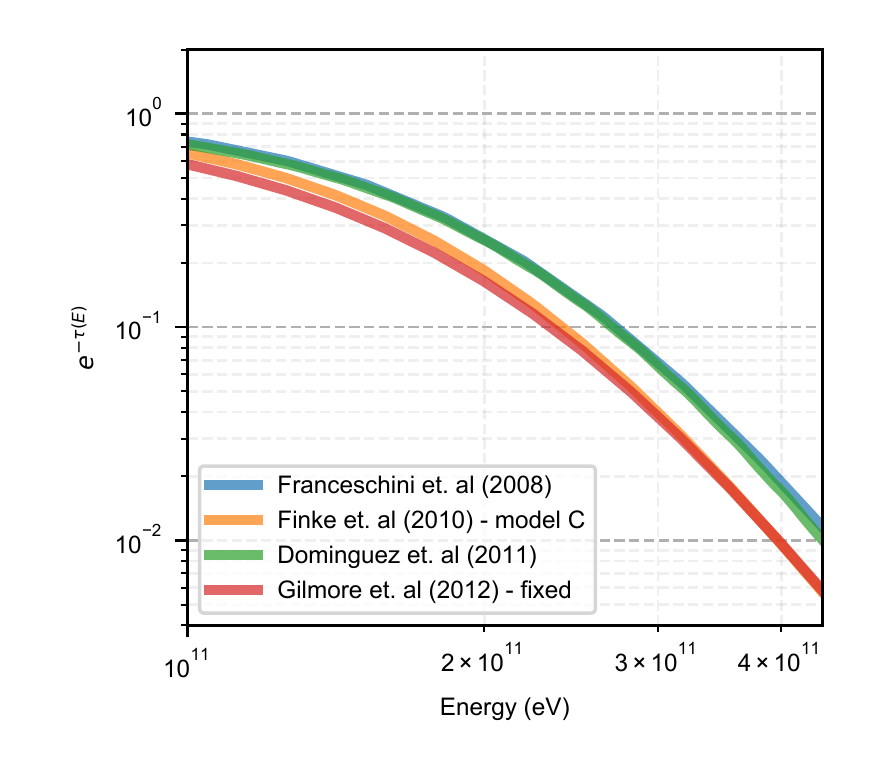}
    \caption{\textbf{EBL absorption coefficient.} Absorption coefficient $e^{-\tau(E)}$ for a source emitting at a redshift 0.653. The values are shown in the energy range of the detected emission of GRB~180720B (100-440 GeV) for the four EBL models considered~\cite{Franceschini2008,Dominguez2011,Finke2010,Gilmore2012}. }
    \label{fig:ebl}
\end{figure}

\begin{figure}[H]
    \centering
    \includegraphics[width = 0.5\textwidth]{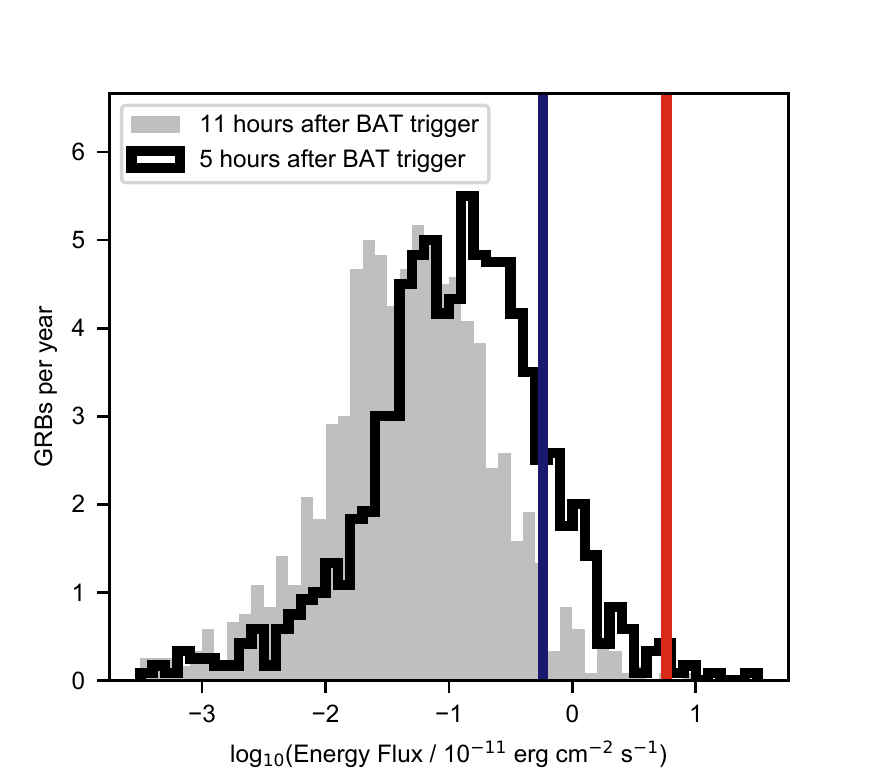}
    \caption{\textbf{CTA detectability prospects.} Energy flux distribution at 11 hours and 5 hours after the \Swift-BAT trigger for all the GRBs observed by the \Swift-XRT per year. The blue vertical line shows the expected sensitivity of CTA, assuming the detectability of fluxes 10 times fainter than GRB~180720B. The energy flux of GRB~180720B is indicated by the red vertical line.}
    \label{fig:CTAprosp}
\end{figure}

\begin{table}[ht!]
 \refstepcounter{table}\label{tab:spectra}
 {\bf Extended Data Table~1~~VHE Spectral information from GRB~180720B}
    \begin{tabular}{llll}
    \hline
    Spectral model &  $F_0$ [$\times$ 10$^{-10}$ cm$^{-2}$ s$^{-1}$ TeV$^{-1}$] & $\photonindex$ & E$_0$ [TeV]   \\
    \hline
    $F_0 \times \frac{E}{E_0}^{-\photonindex}$                        & \,\,\,$2.71\,\pm\,0.74\,^{+1.43}_{-1.16}$      & $3.7\,\pm\,1.0\,^{+0.2}_{-0.1}$ & $0.154$ \\
    $F_0 \times \frac{E}{E_0}^{-\photonindex} \times e^{-\tau(z,E)}$  & \,\,\,$7.52\,\pm\,2.03\,^{+4.53}_{-3.84}$ & $1.6\,\pm\,1.2\,^{+0.4}_{-0.4}$ & $0.154$ \\
    $F_0 \times \frac{E}{E_0}^{-2} \times e^{-\tau(z,E)}$             & $16.12\,\pm\,4.37\,^{+10.59}_{-9.25}$     & $2.0$ [Fixed] & $0.105$ \\
    \hline
    \end{tabular}
    \caption*{Spectral parameters of the fits to the H.E.S.S. observed emission in the energy range 100--440\,GeV.  The intrinsic spectrum with $\photonindex=2.0$ (third row) is provided as a reference to the \Fermi-LAT mean photon index detected in several other GRBs at HE~\cite{FermiGRBCat}. All reported uncertainties are statistical and systematic in that order.}
\end{table}

\end{document}